\documentstyle[prd,aps,floats,epsfig]{revtex}
\tightenlines
\def\be{\begin{equation}}
\def\ee{\end{equation}}
\def\bea{\begin{eqnarray}}
\def\eea{\end{eqnarray}}

\begin{document}
 %@twocolumnfalse\endcsname
\title{Does nonorthogonality necessarily imply nonmaximal entanglement ?}
\author{Dipankar Home$^\dagger$ and M K Samal$^* $}
\address{{\it $^\dagger$ Department of Physics, Bose Institute, Calcutta 
700 009, India. \\
$^*$ S N Bose National Center for Basic Sciences, JD/III,
 Salt Lake, Calcutta 700 098, India.}}
\date{\today}

\maketitle

\begin{abstract}
We show that bipartite entanglements involving non-orthogonal states are
{\it necessarily nonmaximally} entangled, however {\it small} the
non-orthogonality may be! How the deviation from maximal entanglement is
related to nonorthogonality is quantified by using two independent
measures of entanglement corresponding to the violation of Bell's
inequality and the entropy measure respectively. This result is true even
if one of the subsystems has orthogonal states. An application of this is
that a maximal violation of Bell's inequality in entangled neutral kaons
is not possible in the presence of CP violation.
\end{abstract}
%]

\section{Introduction}

Quantum entanglement is usually studied using entangled orthogonal states.  
Although entangled non-orthogonal states play important role in quantum
cryptography\cite{fuchs}, they have received less attention compared to
their orthogonal counterparts, such as in connection with Bell's
inequalities (BI)\cite{sanders1}. Examples of entangled non-orthogonal
states are readily available; entangled coherent states fall under this
category and experiments on quantum teleportation involve
them\cite{sanders2,sanders3}. A Schr\"odinger cat state which has been
developed for $SU(2)$ coherent states could be extended to an entanglement
of nonorthogonal $SU(2)$ coherent states\cite{sanders4}. The entangled
neutral kaons in the presence of CP violation involve such nonorthogonal
entangled states\cite{home88}. Hence a study of the question of how
non-orthogonality is related to maximal entanglement is relevant to the
questions concerning teleportation, dense-coding and entanglement swapping
involving entangled non-orthogonal states. In particular, we may note that
perfect teleportation is possible essentially for maximally entangled
states\cite{bennett} while arguments for nonlocality such as those of the
Hardy-type\cite{hardy} hold for nonmaximally entangled states.

In this paper it is demonstrated that bipartite entangled systems having
non-orthogonal states are {\it necessarily non-maximally} entangled
however {\it small} the non-orthogonality may be, which is true even if
one of the subsystems has orthogonal states. A quantitative relation
between the departure from maximal entanglement and nonorthogonality is
derived by using two independent measures of entanglement corresponding to
the amount of violation of BI and the entropy measure respectively. An
application of this result is discussed for entangled neutral kaons in the
presence of CP violation. We also point out directions for some further
studies. \section{Preliminaries} A bipartite entangled state can, in
general, be written as \be |\Psi>^{AB} = \mu |\alpha>^A |\beta>^B + \nu
|\gamma>^A |\delta>^B, \ee where $|\alpha>^A$ and $|\gamma>^A$ are the
states of system 1 and similarly for $|\beta>^B$ and $|\delta>^B$ for
system 2 with complex coefficients $\mu$ and $\nu$. A bipartite entangled
state involving nonorthogonal states would have the property that the
overlaps $^A<\alpha|\gamma>^A$ and $^B<\delta|\beta>^B$ are non-zero. The
two linearly independent nonorthogonal states $|\alpha>^A$ and
$|\gamma>^A$ that span a two-dimensional subspace of the Hilbert space can
be chosen such that \be |\alpha>^A = \left( \begin{array}{c} 0 \\ 1 \\
\end{array} \right )^A , \:\: |\gamma>^A = \left( \begin{array}{c} {\cal
N}^A \\ y \\ \end{array} \right )^A \ee where ${\cal N}^A = \sqrt{1 -
|y|^2}$ where $y = ^A<\alpha|\gamma>^A$. Similarly for system B one has
\be |\delta>^B = \left( \begin{array}{c} 0 \\ 1 \\ \end{array} \right )^B
, \:\: |\beta>^B = \left( \begin{array}{c} {\cal N}^B \\ x \\ \end{array}
\right )^B \ee where ${\cal N}^B = \sqrt{1 - |x|^2}$ where $x =
^B<\delta|\beta>^B $. In this basis the state (1) can be written as \bea
\Psi^{AB} & =& \mu \left( \begin{array}{c} 0 \\ 1 \\ \end{array} \right
)^A \otimes \left( \begin{array}{c} {\cal N}^B \\ x \\ \end{array} \right
)^B + \nu \\ \nonumber &&\left( \begin{array}{c} {\cal N}^A \\ y \\
\end{array} \right )^A \otimes \left( \begin{array}{c} 0 \\ 1 \\
\end{array} \right )^B = \left( \begin{array}{c} 0 \\ \nu {\cal N}^A \\
\mu {\cal N}^B \\ {\cal M} \\ \end{array} \right ), \eea where ${\cal M}
\equiv \mu x + \nu y$ and the superscripts $A$ and $B$ are ignored for
cases where no ambiguity arises. Also the normalization of (4) requires
that \be | \mu {\cal N}^B|^2 + | \nu {\cal N}^A|^2 + |{\cal M}|^2 =1. \ee

The state (4) is a pure state with density matrix $\rho^{AB}=
\Psi^{AB} {\Psi^{AB}}^\dagger$ and the reduced density matrices
$\rho^A$ and $\rho^B$ for systems $A$ and $B$ are
\begin{eqnarray}
\rho^A &= &{\rm Tr}_B \rho^{AB} =  \left( \begin{array}{cc}  | \nu
{\cal N}^A|^2  &\nu {\cal N}^A {\cal M}^*  \\ \nu^* {\cal N}^A
{\cal M} & | \mu {\cal N}^B|^2 + |{\cal M}|^2  \\ \end{array}
\right ) \nonumber \\ \rho^B &=& {\rm Tr}_A \rho^{AB} =	 \left(
\begin{array}{cc}  | \nu {\cal N}^B|^2	&\nu {\cal N}^B {\cal M}^*
\\ \nu^* {\cal N}^B {\cal M} & | \mu {\cal N}^A|^2 + |{\cal M}|^2
\\ \end{array} \right )
\end{eqnarray}
which satisfy
\be
det \:\: \rho^A = | \mu \nu {\cal N}^B {\cal N}^A|^2 = det \:\;
\rho^B. \ee Both  $\rho^A$ and $\rho^B$ have identical eigenvalues
given by
\be
\lambda_\pm = \frac{1}{2} \pm \frac{1}{2} \sqrt{1 - 4 | \mu \nu
{\cal N}^B {\cal N}^A|^2} \ee with corresponding eigenvectors
$|\pm>^A$ and $|\pm>^B$ for  $\rho^A$ and $\rho^B$ respectively.
Using the general theory of Schmidt decomposition the state
(4) can be expressed as
\be
|\Psi>^{AB} = c_- |->^A |->^B + c_+ |+>^A |+>^B,
\ee
with
\be
|c_\pm|^2 = \lambda_\pm , \:\: |c_-|^2 + |c_+|^2 =1, \ee for 
studying
the violation of BI\cite{sanders2}. Considering each two-state
system as a spin- $\frac{1}{2}$ system, the Hermitian operators
$\hat{\Theta}$ for each system $A$, $B$ are chosen to have the
general form
\bea
\hat{\Theta}&=& \cos{\chi} [\: |+><+| - |-><-|
\:] +
\\ \nonumber && \sin{\chi} [ \: e^{i \phi} |+><-| +
e^{-i \phi} |-><+| \:], \eea
that corresponds to the components of
a `spin' operator along the axis determined by the angles $\chi$
and $\phi$. For the choices \bea \chi^A &=& 0, \:\: \chi^{'A} =
\pi /2 \nonumber\\ \chi^B &=& - \chi^{'B} = \cos^{-1} \: [1 + |2
c_+ c_-|^2]^{- \frac{1}{2}} \nonumber\\ \phi^A + \phi^B & =&
\phi^{'A} + \phi^{'B} = \phi_+ - \phi_- \eea where $\phi_\pm$ are
the phases of $c_\pm$, the expectation value of the Bell operator
\be
\hat{B} = \hat{\Theta}^A  \hat{\Theta}^B +  \hat{\Theta}^A
\hat{\Theta}^{'B} + \hat{\Theta}^{'A}
 \hat{\Theta}^B -  \hat{\Theta}^{'A}  \hat{\Theta}^{'B}
\ee
for the state (9) can be shown to be
\be
B \equiv <\Psi|\hat{B}|\Psi> = 2 \sqrt{1 + |2 c_+ c_-|^2} > 2.
\ee
It was pointed out by Mann et. al. \cite{sanders2} that a violation
of BI always occur. But they did not investigate the question
concerning maximal violation of BI for non-orthogonal entanglements. 

\section{The Violation of BI and Nonorthogonal states}
\subsection{Maximal Violation of BI}

In this subsection we discuss whether or not a maximal violation
of BI is possible when both or one of the pairs of entangled states
is mutually nonorthogonal and the above mentioned Bell operator
is used. We first note from eqn.(14) that the degree of violation
of BI depends on the values of $c_\pm$. 
It is evident that maximal violation occurs
when
\be
|c_+ c_-|^2 = \frac{1}{4} \ee Solving the above equation along
with the normalization condition $ |c_+|^2 + |c_-|^2 = 1$ one
obtains $ |c_+|^2 = |c_-|^2 = \frac{1}{2}$. Substitution of this
in the expression for the eigenvalues of the reduced density
matrices yields
\be
(1 - |x|^2) (1 - |y|^2) |\mu \nu|^2 = \frac{1}{4} \ee The
normalization condition given by eqn.(5) can be written as
\be
|\mu|^2 + |\nu|^2 + 2 |\mu| |\nu| |x| |y| \cos{\eta} = 0, \ee with
$\eta = (\theta_1 - \theta_3 + \theta_2 - \theta_4) $ where
$\theta_j, (j = 1, 2, 3, 4)$ correspond to the phase of $\mu, x,
\nu, y$ respectively. We will show below that for the cases
involving nonorthogonal bases there cannot be a maximal violation
of BI because eqn.(16) and eqn.(17) cannot be simultaneously
solved.

{\bf Case I}: {\it Nonorthogonal-Nonorthogonal(NN) Case}

Substitution of eqn.(16) into eqn.(17) yields in this case
\be
a q^2 + b q + c = 0, \ee where $q = |\mu|^2 $ and the coefficients
$a, b, c$ are given by \bea a & = & 4 (1- |x|^2)(1 - |y|^2)
\nonumber \\b & =& a \big[ \sqrt{\frac{|x|^2 |y|^2}{(1- |x|^2)(1 -
|y|^2)}} \cos{\eta} - 1\big] \nonumber \\ c & = & 1. \eea The positive
definiteness of $q$ requires $b^2 - 4 a c \ge b^2$ when $b > 0$
and $b^2 - 4 a c \le b^2$ when $b < 0$. The first of these
conditions cannot be satisfied for the given $a, b$ and $c$. But
the second condition is satisfied when \be \cos{\eta} \le
\frac{\sqrt{(1 - |x|^2)(1 - |y|^2)}}{|x||y|}. \ee Since $q$ has to
be real one has $b^2 \ge 4 a c$ which in turn implies that \be
\cos{\eta} \ge \frac{1 + \sqrt{(1- |x|^2)(1 - |y|^2)}}{|x||y|},
\ee which is clearly in conflict with eqn.(20). Since $-1 \le
\cos{\eta} \le 1$ the above condition requires
\be
\frac{1 + \sqrt{(1- |x|^2)(1 - |y|^2)}}{|x||y|} \le -1 \ee which
cannot be satisfied. Hence $q$ cannot be real which means that
simultaneous solution of eqn.(16) and eqn.(17) is not possible.
Thus it is not possible to have maximal violation of BI in this
case.

{\bf Case II}:{\it Orthogonal-Nonorthogonal(ON) Case}

In this case, bases for the system $A$ is chosen to be orthogonal
which means $|y| = 0$ and ${\cal N}^A = 1$. The coefficients of
the quadratic equation involving $q$ are given by \be a  =  4
(1- |x|^2); \: b  = - a; \:  c  =  1. \ee
Although the positive definiteness of $q$ may be satisfied in this
case, $q$ cannot be real. Because for $q$ to be real one should
have $(1- |x|^2) \ge 1$ which is not possible. Hence we find that
even if one of the bases is orthogonal still then it is not
possible to have maximal violation of BI in an entangled bipartite
system.

{\bf Case III}:{\it Orthogonal-Orthogonal(OO) Case}

The orthogonality for both bases requires $|y|= 0$ and $|x|=0$.
The coefficients of the quadratic equation involving $q$ are given
by \be a = 4; \: b = - a; \:  c = 1.
\ee It is evident that there is a maximal violation of BI in this
case because $q = \frac{1}{2}$ is a solution which in turn means
$|\mu|^2=|\nu|^2=\frac{1}{2}$.

\subsection{Non-Maximal Violation of BI}

Now we discuss the explicit relationship between the parameters
$x$ and $y$ (that represent non-orthogonality of the bases) and
the parameter $d$ that denotes the deviation from the maximal
violation of BI. We define the parameter $d$ by the following
equation:
\be
\sqrt{1 + |2 c_+ c_-|^2} = \sqrt{2 - d} \ee where $0 \le d \le 1$
and the lower limit corresponds to maximal violation of BI. In
this case eqn.(16) becomes a function of $d$ as follows:
\be
(1 - |x|^2) (1 - |y|^2) |\mu \nu|^2 = \frac{1-d}{4} \ee
Substituting the above eqn. into eqn. (17) yields the quadratic
equation for the {\it NN} case
\be
q^2 + b q + c = 0, \ee where $q = |\mu|^2 $ and the coefficients
$b, c$ are given by \bea b & =& [ \sqrt{\frac{|x|^2 |y|^2
(1-d)}{(1- |x|^2)(1 - |y|^2)}} \cos{\eta} - 1] \nonumber \\ c
& = & \frac{(1-d)}{4 (1- |x|^2)(1 - |y|^2)}. \eea The most general
solution to eqn.(27) is \bea |\mu|^2 &=& \frac{1}{2}[(1- X
\cos{\eta}) \nonumber \\ && \pm \sqrt{(1- X \cos{\eta})^2
- \frac{X^2}{|x|^2 |y|^2}}], \eea where
\be
X =\sqrt{\frac{|x|^2 |y|^2 (1-d)}{(1- |x|^2)(1 - |y|^2)}} \ge 0.
\ee This solution reduces to maximal violation of BI in various
cases discussed earlier for the case $d = 0$. Similar relations
for the {\it ON} case and {\it OO} case respectively are given by
\bea |\mu|^2 & = & \frac{1}{2} \pm \frac{1}{2} \sqrt{\frac{d -
|x|^2}{1 - |x|^2}}
\\ \nonumber |\mu|^2 & = & \frac{1}{2} \pm \frac{1}{2} \sqrt{d}.
\eea 

It is easy to see from the above expressions that there is no
maximal violation of BI in {\it ON} case whereas there is a
maximal violation of BI in {\it OO} case. Since the maximal
violation of BI provides a measure of the degree of entanglement
we can now indicate
explicitly how the degree of entanglement (denoted by the
deviation parameter $d$) changes with the overlap of the
nonorthogonal bases:
\bea
d & =& 1 - 4 (1- |x|^2)(1 - |y|^2)[|\mu|^2 + \\ \nonumber & & 
|\mu|^4 (2 |x|^2 |y|^2
\cos^2{\eta} - 1) \pm \sqrt{2} Z |\mu|^3 |x| |y| 
\cos{\eta}],
\eea
where
\be
Z = \sqrt{2 (1 - |\mu|^2) + |\mu|^2 |x|^2 |y|^2 (1 + \cos{2 \eta})}.
\ee
The above relation reduces to
\bea
d & = & 1 - 4 |\mu|^2 (1 - |\mu|^2) (1 - |y|^2) \\ \nonumber
d & = & 1 - 4 |\mu|^2 (1 - |\mu|^2). 
\eea
for the {\it ON} case and {\it OO} case respectively.
It is evident from the above relations that for the {\it OO} case
$d$ vanishes, thereby implying a maximal violation of BI.

\section{Measure of Entanglement and Nonorthogonal States}

Another way of seeing how the nonorthogonality of the bases can
affect the measure of entanglement is to note that the von Neumann
entropy measure of entanglement of bipartite state $\Psi^{AB}$ is
given by
\be
{\cal E}(\psi^{AB}) = - {\rm Tr} (\rho^A \ln \rho^A) = - {\rm Tr}
(\rho^B \ln \rho^B),\ee where the reduced density matrices
$\rho^A$ and $\rho^B$ are given as a function of the parameters
$x$ and $y$ through eqn.(6). The above expression can be
written as\cite{woo} \bea {\cal E}(\psi^{AB}) &= &h(\frac{1 +
\sqrt{1 - C^2}}{2});\\ \nonumber h(Z) & = & - Z {\rm log}_2 Z - (1
- Z) {\rm log}_2 (1- Z ), \eea where the {\it concurrence} $C$ is
defined as
\be
C(\Psi) = |<\Psi|\tilde{\Psi}>| \ee The $|\tilde{\Psi}>$ that
appears in the definition of $C$ is given by
\be
|\tilde{\Psi}> = \sigma_y |\Psi^*>, \ee where $|\Psi^*>$ is the
complex conjugate of $|\Psi>$  when it is expressed in a fixed
basis such as $\{|\uparrow>, |\downarrow>\}$, and $\sigma_y$
expressed in that same basis is the usual Pauli matrix. For a
spin-$\frac{1}{2}$ particle this is the standard time reversal
operation and it reverses the direction of the spin.

In our case, we have \bea C^2(\Psi^{AB}) &=& 4 {\rm det}\rho^A = 4
{\rm det}\rho^B \\ \nonumber & = & 4 (1 - |x|^2) (1 - |y|^2) |\mu
\nu|^2. \eea 
It is easy to see from the eqn.(39) and eqn. (16) that 
the {\it concurrence} $C$
becomes $1$ for the maximally entangled states. 
For any non-maximally entangled state it is always less than $1$.
Thus eqn.(39) and eqn. (36) indicate how the entropy measure of the
entanglement depends on the nonorthogonality of the bases.

\section{Applications}
\subsection{Neutral Kaons System}

As an application of the formalism developed above we show here
the relationship between the deviation from maximal violation of
BI (parameter $d$) and the {\it CP} violation parameter $\epsilon$
for neutral kaons. For them the
mass eigenstates $|K_S>$ and $|K_L>$ are written in terms of {\it
CP} eigenstates $|K_1>$ and $|K_2>$ as \bea |K_S> & = &
\frac{1}{\sqrt{1 + |\epsilon|^2}} [ |K_1> + \epsilon |K_2>] \\
\nonumber |K_L> & = & \frac{1}{\sqrt{1 + |\epsilon|^2}} [ |K_2> +
\epsilon |K_1>], \eea where \be |K_{1, 2}> = \frac{1}{\sqrt{2}} [
|K^0> \pm \bar{|K^0>}] \ee In this system, the nonorthogonal
overlaps are given by
\be
|x|^2 = |y|^2 = |<K_S|K_L>|^2 = (\frac{{\it Re} \epsilon}{1 +
|\epsilon|^2})^2 \ee 

Taking into account the charge conjugation ($= - 1$) of the 
$\Phi$--meson the
EPR-Bohm type entangled state of the neutral kaon pair coming from 
$\Phi$ decay can be written as
\begin{equation} \label{1}
|\Phi \rangle = {1 \over \sqrt{2}}\left[|K^0\rangle \otimes|\bar{K^0}\rangle
-|\bar{K^0}\rangle \otimes|K^0 \rangle \right].
\end{equation}
The neutral kaons then fly apart getting spatially separated. 
Their time evolution under weak interactions is given by
\be
|\Phi \rangle = \frac{N(t)}{\sqrt{2}} \left[|K_S\rangle 
\otimes|K_L\rangle - |K_L \rangle \otimes|K_S \rangle \right], 
\ee
where $|N(t)| = (1 + |\epsilon|^2)/(|1 - \epsilon^2|) 
\times e^{- \frac{1}{2} (\Gamma_S + \Gamma_L) t }$ reflecting
the extinction of the beams via weak interaction induced 
kaon decays but without
modifying the perfect antisymmetry of the initial state.

For this system the deviation parameter $d$ is given as a 
function of $\epsilon$
as follows:
\be
d = 1 - \{1 - (\frac{{\it Re} \epsilon}{1 +
|\epsilon|^2})^2 \}^2 [1 + \sqrt{2} Y \cos{\eta} 
({\it Re} \epsilon)^2 (1 +
|\epsilon|^2)^{-4}], 
\ee
where
\be
Y = \sqrt{2} \cos{\eta} 
({\it Re} \epsilon)^2 \pm \sqrt{({\it Re} \epsilon)^4 + 
({\it Re} \epsilon)^4 \cos{2 \eta} + 2 (1 +
|\epsilon|^2)^4}. 
\ee

Hence from eqn.(29) it is clear that there cannot be a maximal violation
of BI for entangled neutral kaons in the presence of {\it CP} violation.
But the deviation $d$ can be made quite small compared to $1$ by suitable
choice of $\eta$ since CP violation parameter $\epsilon \approx 10^{-3}$.

\section{Conclusion}

To summarise, we have shown that the bipartite entangled systems involving
non-orthogonal states are {\it non-maximally} entangled, irrespective of
how {\it small} the non-orthogonality is. Now, note that the physical
content of entanglement (e. g., the possibility of perfect teleportation,
maximal violation of local realism) is contingent on the degree of
entanglement and in particular on whether entanglement is maximal.
However, any measure of entanglement changes under a {\it non-unitary}
transformation that connects a non-orthogonal basis with an orthogonal
basis. Thus our work implies that, no matter how small this non-unitarity
is, such a transformation {\it changes} the physical content of
entanglement.  Implications of this feature could be interesting for
further studies.

One application of the above result is pointed out by showing that a
maximal violation of BI in entangled neutral kaons is not possible in the
presence of CP violation. Similar applications can be made in other
examples as well, such as for entangled coherent states\cite{sanders2}. It
may also be noted that in an experiment using orthogonal entangled states,
any finite imprecision in the preparation of the entanglement can result
in a departure from the required orthogonality. In order to analyse how
such a deviation from the orthogonality in the prepared entangled state
affects the relevant experimental result, the results presented in our
paper could be useful.

\section{Acknowledgement}  

This research is funded by the Dept. of Science and Technology, Govt. of
India. We acknowledge the helpful discussions with the participants of the
First Winter School on Foundations of Quantum Physics and Quantum Optics,
Calcutta (Jan., 2000).

\end{document}